\documentclass[twocolumn,prx, aps,superscriptaddress,showpacs]{revtex4-2}
\usepackage{subfigure}
\usepackage{psfrag}
\usepackage{dcolumn}
\usepackage{bm}
\usepackage{color}
\usepackage{latexsym}
\usepackage{epstopdf}
\usepackage[english]{babel}
\usepackage{times}
\usepackage{latexsym}
\usepackage{natbib}
\usepackage{graphicx}
\usepackage{epsf}
\usepackage{amsmath}
\usepackage{amssymb}
\usepackage{amsfonts}
\usepackage{appendix}
\usepackage{lipsum, babel}
\usepackage{soul}
\usepackage{hyperref}
\hypersetup{colorlinks=true,citecolor=myblue,linkcolor=blue,urlcolor=myblue}

\usepackage{siunitx}

\definecolor{mygrey}{gray}{0.35}
\definecolor{myblue}{rgb}{0.2,0.2,0.8}
\definecolor{myzard}{cmyk}{0,0,0.05,0}
\definecolor{mywhite}{rgb}{1,1,1}
\definecolor{myred}{rgb}{1,0.,0.3}
\definecolor{myblack}{rgb}{0,0,0}

 \newcommand{\ket}[1]{|#1\rangle}

\newcommand{\up}{\mid\uparrow\uparrow\rangle}

\usepackage{dsfont}

\makeatletter

\renewenvironment{widetext@grid}{%
  \par\ignorespaces
  \setbox\widetext@top\vbox{%
   \vskip15\p@
   \hb@xt@\hsize{%
    \leaders\hrule\hfil
    \vrule\@height6\p@
   }%
   \vskip6\p@
  }%
  \setbox\widetext@bot\hb@xt@\hsize{%
    \vrule\@depth6\p@
    \leaders\hrule\hfil
  }%
  \onecolumngrid
  \let\set@footnotewidth\set@footnotewidth@ii
}{%
  \par
  \twocolumngrid\global\@ignoretrue
  \@endpetrue
}%

\makeatother
\begin{document}
\raggedbottom
\title{The effect of fast noise on the fidelity of trapped-ions quantum gates }
\author{Haim Nakav}
\thanks{These authors contributed equally to this work}

\affiliation{Physics of Complex Systems, Weizmann Institute of Science and AMOS, Rehovot 7610001, Israel}
\author{Ran Finkelstein}
\thanks{These authors contributed equally to this work}

\affiliation{Physics of Complex Systems, Weizmann Institute of Science and AMOS, Rehovot 7610001, Israel}
\altaffiliation{ Current address : Division of Physics, Mathematics and Astronomy, California Institute of Technology, Pasadena, CA 91125, USA}

\author{Lee Peleg}
\affiliation{Physics of Complex Systems, Weizmann Institute of Science and AMOS, Rehovot 7610001, Israel}
\author{Nitzan Akerman}
\affiliation{Physics of Complex Systems, Weizmann Institute of Science and AMOS, Rehovot 7610001, Israel}
\author{Roee Ozeri}
\affiliation{Physics of Complex Systems, Weizmann Institute of Science and AMOS, Rehovot 7610001, Israel}
\begin{abstract}
High fidelity single and multi-qubit operations compose the backbone of quantum information processing. This fidelity is based on the ability to couple single- or  two-qubit levels in an extremely coherent and precise manner. A necessary condition for coherent quantum evolution is a highly stable local oscillator driving these transitions. Here we study the effect of fast noise, that is noise at frequencies much higher than the local oscillator linewidth, on the fidelity of one- and two-qubit gates in a trapped-ion system. We analyze and measure the effect of fast noise on single qubit operations including resonant $\pi$ rotations and off-resonant sideband transitions . We further analyze the effect of fast phase noise on the M\o lmer-S\o rensen two-qubit gate. We find a unified and simple way to estimate the performance of all of these operations through a single parameter given by the noise power spectral density at the qubit response frequency. While our analysis focuses on phase noise and on trapped-ion systems, it is relevant for other sources of fast noise as well as for other qubit systems in which spin-like qubits are coupled by a common bosonic field. Our analysis can help in guiding the deign of quantum hardware platforms and gates, improving their fidelity towards fault-tolerant quantum computing.

\end{abstract}
\maketitle
\section{Introduction}
Quantum coherence is the fundamental resource in any quantum information processing task. Much of the remarkable progress achieved in the field over the past two decades was achieved by careful examination and reduction of dephasing mechanisms through improvements in control technology, and mitigation of environmental noise \cite{benhelm2008towards,schindler2013quantum}. Yet, single and two-qubit gates are still often limited by the finite coherence between the qubit and the classical local oscillator (LO) used to control and measure the quantum system.


In quantum information processing, phase noise typically limits the coherence time of quantum registers and compromises the fidelity of quantum gates. Many studies investigated this effect, and techniques to mitigate it were proposed and implemented. These include dynamical decoupling \cite{ban1998photon,PhysRevLett.82.2417,PhysRevA.59.4178,PhysRevA.58.2733,PhysRev.94.630,meiboom1958modified,PhysRevLett.119.220505}, and decoherence-free subspaces \cite{lidar1998decoherence,barenco1997stabilization,PhysRevA.63.012301,PhysRevLett.79.1953,PhysRevLett.79.3306}. These methods were designed to mitigate the effects of slow phase noise ($\sigma_z$ errors), which refers to noise within the linewidth of the oscillator as it acts on the qubit transition. However, noise that is faster than this linewidth affects qubit performance differently. In some cases, which will be discussed in this paper, such noise acts as a bit flip error, rather than a $\sigma_z$ error, rendering the above-mentioned methods ineffective.
\par

Fast LO noise can result from various mechanisms. One example of such a source is a side effect of reducing the magnitude of slow phase noise. In order to narrow the linewidth of an oscillator often an external servo loop is added to suppress the slow phase noise of the LO with respect to a stable reference. While subduing the slow phase noise, the servo loop generates excess noise at the higher frequencies close to its unity gain response \cite{ogata2010modern}. This noise feature, which is prevalent mainly in narrow linewidth lasers, is sometimes referred to as servo bump and results in fast phase noise.  \par

Traditionally, noise at these high frequencies was thought to average out over the much longer time scale of typical quantum evolution. 
However, it was recently realized that such noise limits various quantum operations  \cite{Akerman_2015,PhysRevLett.110.110503,fanciulli2009electron,PhysRevLett.103.220802,PhysRevB.77.174509,LASIC2006208,Day2022} if it overlaps with a frequency at which the quantum system resonantly responds.\par

\begin{figure}[tb!]
    \centering
    \includegraphics[width=\columnwidth]{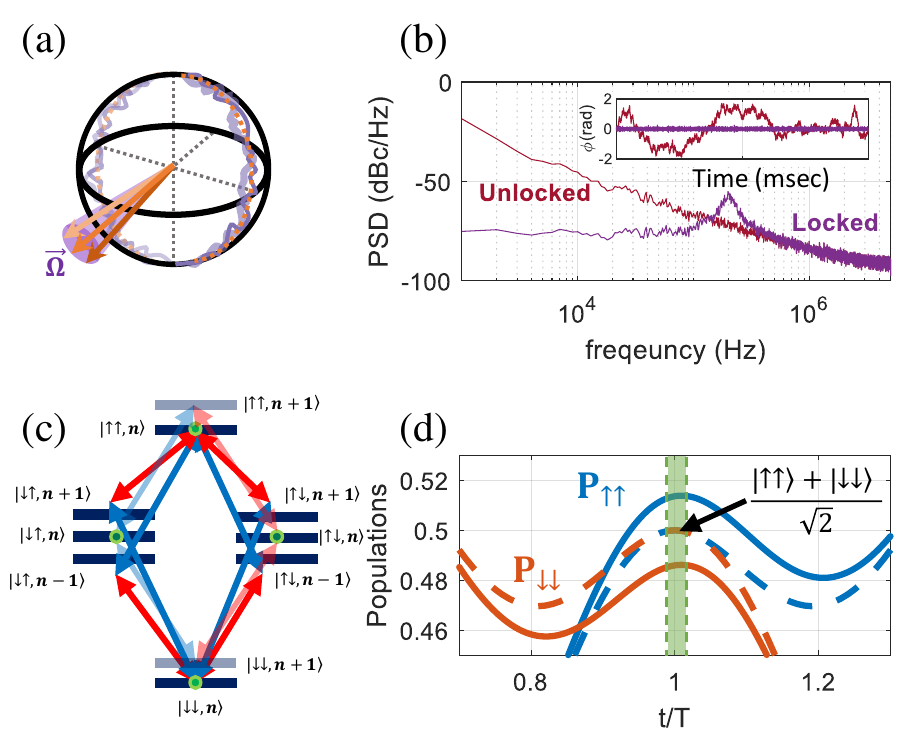}
    \caption{\textbf{The effect of fast noise on the fidelity of trapped-ions quantum gates.} (a) Illustration of the Bloch sphere of a qubit where the Rabi vector orientation fluctuates due to fast phase noise. The orange dashed line is the noise-free trajectory for this qubit operation, whereas the purple lines are trajectories of the qubit in the presence of fast noise. (b) Typical laser phase power spectral densities (PSD). The red curve represents a  typical PSD for a free-running laser which is characterized by a white frequency noise or, equivalently, $1/f^2$ phase noise. The purple curve represents the PSD of the same noise process after applying an external feedback loop that suppresses the slow noise while increasing the noise at the edge of its bandwidth. The equivalent time traces are shown in the inset. (c) M\o lmer-S\o rensen (MS) gate in the presence of fast noise can result in incoherent coupling to other levels resulting in leakage of population to other motional states, decreasing the gate fidelity. (d) The MS interaction dynamics, visualized through the two populations without phase noise (dashed lines) and with noise (solid lines). At gate time (green area), we expect to obtain a Bell state, due to phase noise the qubits reach a slightly erroneous state which includes an incoherent mixture of Bell states.}
    \label{fig:intro}
\end{figure}

As an example, we consider a single qubit rotation on the Bloch sphere shown in Fig. \ref{fig:intro}(a). Phase or frequency instabilities lead to fast fluctuations of the Rabi vector in the Bloch sphere equatorial plane, resulting in randomly modified trajectories [purple trajectories on the Bloch sphere in Fig. \ref{fig:intro}(a)]. Specifically, the fast noise frequency components that will not average out and lead to significant rotation errors are those that overlap the Rabi frequency. Fig \ref{fig:intro}(b) showcases an example of such a modified spectrum.  In fact, the effect of fast noise goes beyond impacting single qubit rotations and can have deleterious effects on the fidelity of two-qubit gates. For example, Starting from $\up$, the M\o lmer-S\o rensen (MS) gate shown in Fig. \ref{fig:intro}(c,d) under fully coherent evolution forms a perfect Bell-state. However, the presence of fast noise that overlaps the carrier transition leads to incoherent errors and reduced fidelity. \par
Here, we simulate and measure the effect of fast noise on the fidelity of single- and two-qubit gates in trapped ion qubits. Single-qubit rotations are performed via resonant pulses and their fidelity was directly measured for different phase-noise spectra. Two-qubit entanglement is performed using the MS gate that uses electromagnetic fields that are close to resonance with the sideband of a common ion-phonon mode. 
We find that the magnitude of the error induced by fast phase noise in quantum gates predominantly scales with the overlap of the noise power spectral density (PSD) with the relevant response frequency. That is, the Rabi frequency in the case of single qubit gates and the detuning from carrier transition in the case of sideband transitions and MS gates. While this study focuses on the effect of phase noise on trapped-ion qubit gates, our results are broadly relevant for any fast noise sources, such as amplitude noise as well. Our results are also relevant for other quantum computing technologies in which two-qubit gates are realized through coupling to a common bosonic mode, such as in superconducting qubits coupled through a microwave resonator.   

 \section{Experimental and numerical setup}
 We perform stochastic master equation simulations and controlled experiments to evaluate the effect of fast noise on the fidelity of quantum operations. Specifically, here we focus on the PSD typical of oscillators where fast noise is amplified by a frequency-stabilization feedback. For spectral frequencies far from the carrier typical oscillators exhibit a white frequency noise spectrum or equivalently, a brown phase noise spectrum $S(\omega)=S_0/\omega^2$ \cite{inbook}. To obtain a random vector with such a PSD and a target root-mean square (RMS) we use a vector of Gaussian distributed independent samples as a seed for a generalized Gauss-Markov stochastic model \cite{langbein2004noise,xu2019easy}. This noise is then convoluted with the impulse response of a phase-locked loop system. The gain and positions of the system's poles can be tuned to shape the spectral response and the resulting output PSD. This results in a spectral region of amplified noise or servo-bump, around the unity gain frequency of the simulated system as shown in Fig. \ref{fig:intro}(b). The random time-series generated through this process are used for both the experiments and the simulations presented below.
 For the simulations, we numerically solve the master equation with a time-dependent phase in the $\sigma_x$ drive term. In each iteration we sample a different random phase vector, while all vectors are drawn from the same PSD. The numerical simulations are performed using the QuTiP\cite{johansson2012qutip} python package. 
 For the different configurations described below, we use different expectation values obtained from the time-evolution of the full density matrix. These values are averaged over the different noise realizations to obtain the results presented here.

\par For the experiment, we use an external cavity diode laser stabilized to a high-finesse cavity. We further utilize the cavity as an extremely narrow filter and take light transmitted through it to perform injection locking on an additional diode, yielding the desired power with low noise. We synthesize phase noise using an arbitrary waveform generator and inject it to the experiment through modulation with an acousto-optical modulator (AOM). The laser is then used to drive the optical qubit transition $\ket{S_{1/2};m_j{=}1/2}-\ket{D_{5/2};m_j{=}3/2}$ in a single $^{88}\mathrm{Sr}^+$ ion trapped in a linear Paul trap.

\section{Single qubit gates}

\begin{figure}[tb!]
    \centering
    \includegraphics[width=\columnwidth]{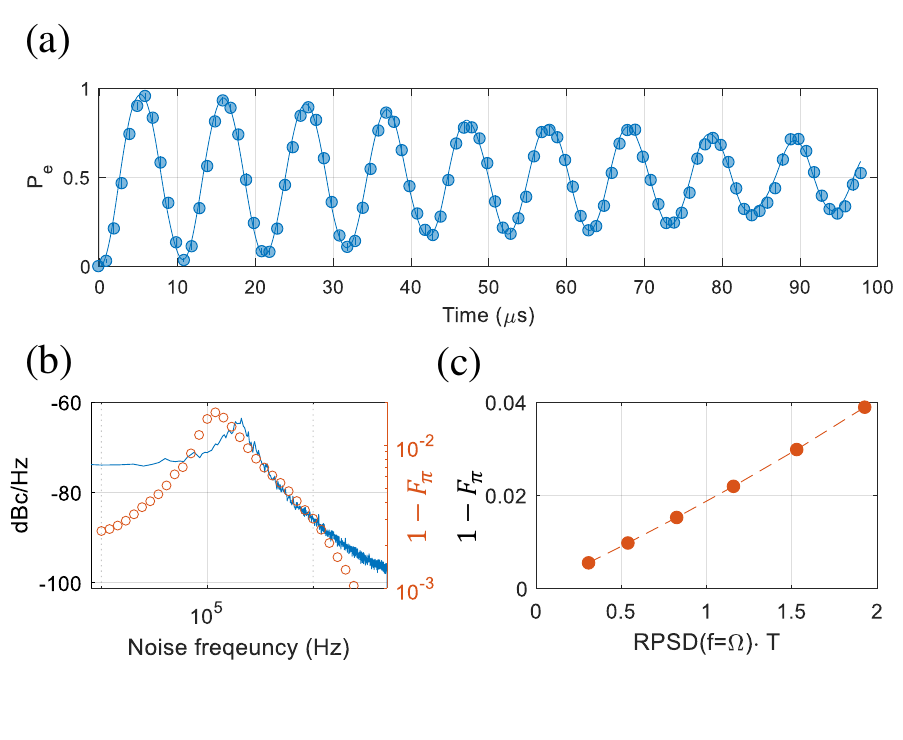}
    \caption{\textbf{Single qubit rotation infidelity due to fast phase noise.} (a) Simulation (solid line) and measurements (filled circles) of Rabi oscillations for an ion driven by a noisy oscillator. Experimental error bars are smaller than the marker size. (b) The infidelity of a single qubit $\pi$ rotation calculated at different Rabi frequencies (orange markers), overlaid on the phase noise PSD characterizing the simulated driving laser. (c)  The infidelity of a single qubit $\pi$ rotation at a fixed Rabi frequency as a function of the Rabi PSD at the Rabi frequency multiplied by the gate time.}
    \label{fig:Fig 2}
\end{figure}

We begin by studying the most simple case of single-qubit rotations in the presence of fast phase noise.
In Fig. \ref{fig:Fig 2}(a) we show the results of continuously driving the qubit transition with a Rabi frequency $\Omega=2\pi\times100 $ kHz and an additional synthesized phase noise characterized by a peak in the PSD around 200 KHz. The experimental measured excited state population (circles) exhibits Rabi oscillations with a decoherence rate which agrees well with the stochastic numerical simulation (solid line) accounting for the the same noise PSD \cite{Peleg2019}. 
We further study numerically the $\pi$-pulse error for a wide range of noise spectra and find that the gate error is linearly dependent on the noise PSD at the Rabi frequency. The relevant PSD in this case is the PSD of the electric field (rather than the phase) driving the transition, $\mathrm{E} = \mathrm{E_0}\cos[\omega t + \phi(t)]$, where $\phi(t)$ is the time-dependent phase which includes the noise term. To evaluate the field PSD in units of Rabi frequency we normalize the entire spectrum such that the area under the carrier peak is $\Omega ^ 2$. We term the new spectral density as the Rabi PSD (RSPD). The dependence of a $\pi$-pulse error vs. the RPSD is shown in Fig. \ref{fig:Fig 2}(c) where we see a linear dependence. For pulses that are long as compared with the inverse of relevant noise features, this dependence makes sense, as the RPSD at the Rabi frequency generates an effective coupling between the Rabi-dresses states. This effect was widely studied in the context of noise spectroscopy \cite{Kotler2011, PhysRevLett.110.110503, DegenRMP2017}.  

For short pulses however, such as $\pi$ rotations, we find a slightly different picture. In Fig. \ref{fig:Fig 2}(b) we plot the noise PSD and the calculated $\pi$ pulse infidelity $1-F_\pi$ for different Rabi frequencies. We find that the largest infidelity does not occur when the Rabi frequency exactly overlaps the peak in the PSD, but rather at lower Rabi frequencies. Considering the $\pi$ pulse length $t_\pi=\pi/\Omega$ we find that the effective bandwidth of such a short pulse is on the order of the Rabi frequency and thus samples, in practice, the entire relevant region of the noise PSD equally. However, the pulse length is still inversely proportional to the Rabi frequency such that longer pulses sample the noise for a longer time. This combined integrated response leads to a shift in the maximal infidelity towards lower Rabi frequencies \cite{Chen2021}. The sensitivity of short gate fidelity to noise integrated over a wide spectral window, renders the proportionality factor between the gate error and the RPSD at the Rabi frequency, seen in Fig. \ref{fig:Fig 2}(c) to be on the order of $10^{-2}$, dependent on the details of the noise spectrum.  

\section{Two qubit gates}

\emph{Sideband transitions.--} Here we consider the M\o lmer-S\o rensen gate \cite{sorensen1999quantum,sorensen2000entanglement} which is a composite operation in which the spin and motion of trapped ions are entangled during the gate through driving of motional sideband transitions. We thus start by disassembling this operation into its primitive constituents. We first study numerically and experimentally the interplay of fast noise and coherent driving of sideband transitions in a single trapped ion. Beyond their role in the MS gate, off-resonant coupling fields are widely used in experiments with atoms and molecules. Such fields are used in generating dressed states, trapping, cooling and local addressing of atomic qubits. It is thus of general relevance to understand the effect of fast noise on sideband transitions. 

\begin{figure}[tb!]
    \centering
    \includegraphics[width=\columnwidth]{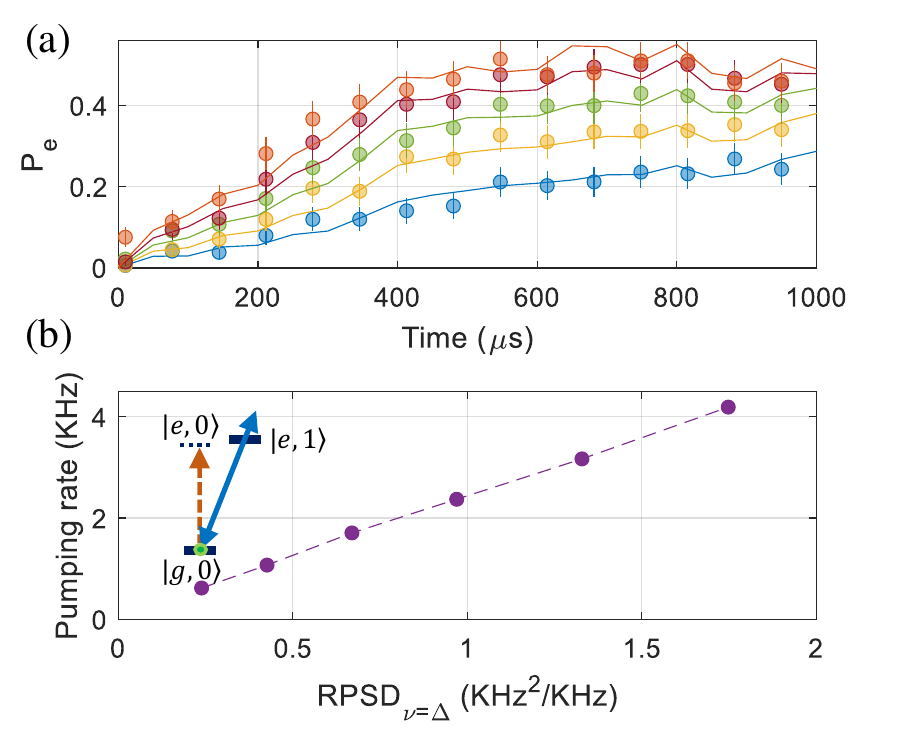}
    \caption{\textbf{Incoherent pumping due to carrier coupling.} (a) Simulation (solid line) and measurements (filled circles) of fraction of atoms in the excited state as a function of off-resonant drive time for drives with different power spectral densities. (b) Incoherent pumping rate as a function of the Rabi frequency power spectral density at the detuning from the carrier transition.}
    \label{fig:side_band_pump}
\end{figure}

We begin by studying the coupling of noise to the spin degree of freedom. We effectively remove any coupling to motional degrees of freedom by detuning the laser central frequency by a few hundreds of KHz from both the carrier transition as well as from any sideband transition. The synthesized noise PSD here has a broad peak around typical trap frequencies $\nu \approx 700$ KHz. Following a drive of variable time, we image the ion to determine its spin state. In Fig. \ref{fig:side_band_pump} we plot the excited spin state population as a function of drive time for different levels of synthesized noise, each averaged over 31 realizations. We observe incoherent spin pumping due to fast noise overlapping the carrier transition. We find good agreement with the numerical simulation, which for a sufficient number of realizations converges well to a simple pumping rate model $\mathrm{P_e}=0.5[1-\mathrm{exp}(-\Gamma t)]$.
 In Fig. \ref{fig:side_band_pump}(b) we plot the pumping rate $\Gamma$ as a function of the RPSD at the detuning frequency from the carrier transition $\Delta$. We find that the pumping rate is proportional to the RPSD at the carrier transition frequency, \emph{i.e.} $\Gamma\simeq 2\cdot \mathrm{PSD}(f=\Delta)$.

\begin{figure}[tb!]
    \centering
    \includegraphics[width=\columnwidth]{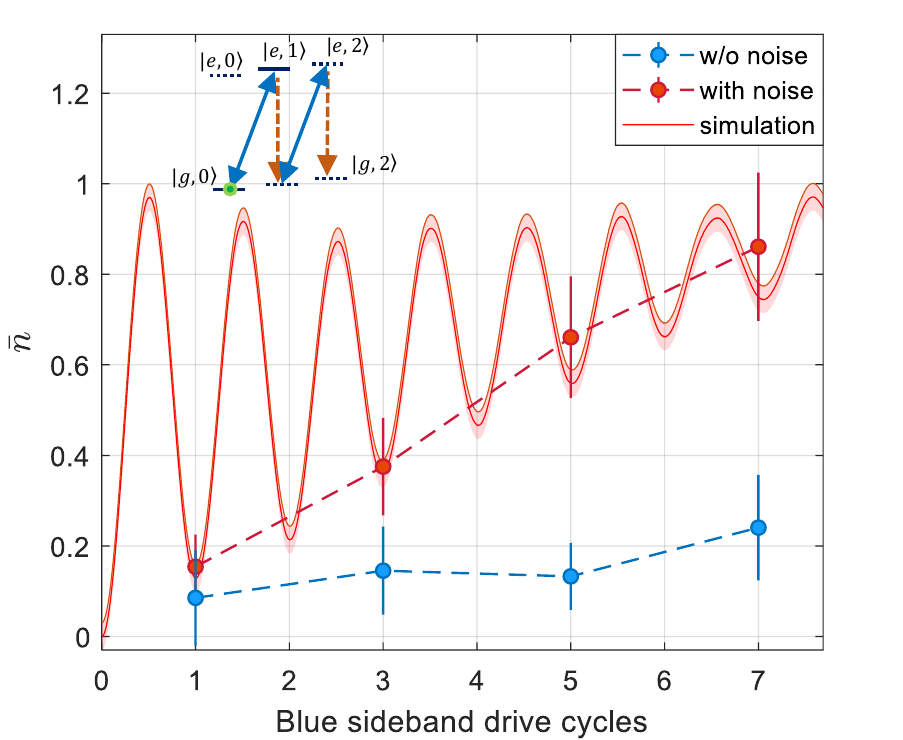}
    \caption{\textbf{Heating due to fast noise in driven sideband transitions.} In the main panel we plot the average number of occupied motional states $\Bar{n}$ inferred from thermometry following an integer number of blue sideband driving cycles. The red markers show measurements following evolution with added synthesized noise while the blue markers show measurements with no added noise (dashed line is a guide to the eye). The inset shows the relevant level scheme where unwanted carrier transitions (orange dashed arrows) are induced due to fast noise.}
    \label{fig:heating}
\end{figure}

In a second experiment, we effectively trace out the spin degree of freedom and measure the occupation of the motional modes. We drive the blue sideband transition for an integer number of spin cycles, and then perform thermometry by driving the blue sideband transition again with a noise-free laser and infer the occupation of motional states from sideband Rabi thermometry \cite{meekhof1996,cai2021}. In Fig \ref{fig:heating}, we plot the measured average number of phonons  $\Bar{n}$ following an integer number of blue sideband cycles. We find excess heating in the presence of fast noise, which grows as a function of the drive time. As a reference, we repeat the measurement in the absence of synthesized noise and find negligible amount of residual heating, which may be attributed to inherent noise in our laser. 

This heating effect is another outcome of the incoherent carrier coupling in the presence of  fast phase noise and driving sideband transition. As seen in the inset of Fig. \ref{fig:heating}, fast noise drives population incoherently on the carrier transition which results in transfer of the ion to higher phonon states. Our numerical simulation in this regime (solid line in Fig. \ref{fig:heating}) reveals the interplay between coherent sideband driving and incoherent population of vibrational modes. As a result, the contrast of coherent Rabi oscillations is reduced along with a constant increase in the average number of phonons. When this number reaches $\bar{n}=1$, the distribution becomes thermal, and we can assign an effective temperature to the ion. Our analysis shows that further driving of the ion would lead to a linear increase in temperature, which we expect to saturate when the motional spread of the ion exceeds the limit of the Lamb-Dicke regime such that carrier and sideband excitations are suppressed.

\begin{figure}[htb!]
    \centering
    \includegraphics[width=\columnwidth]{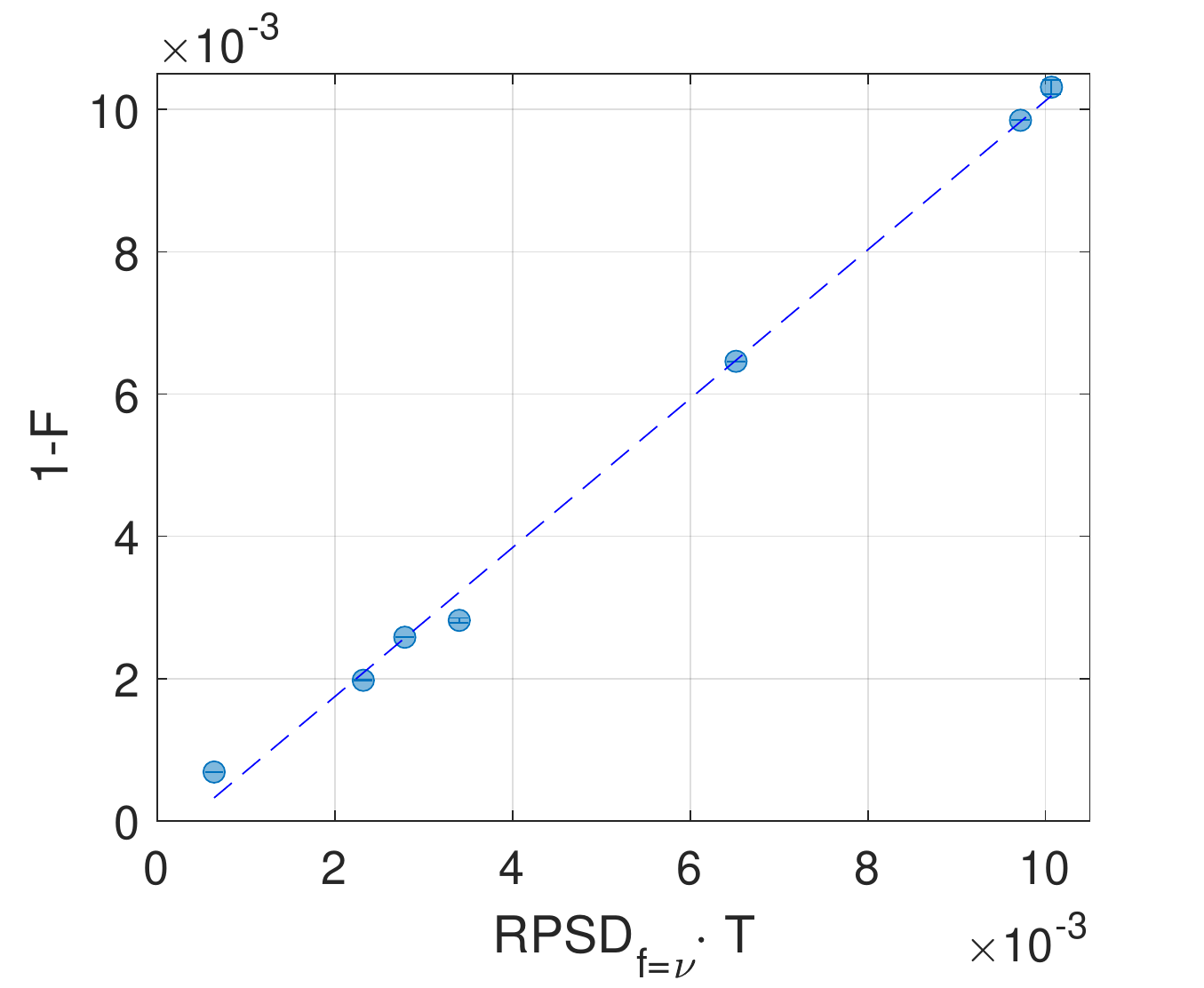}
    \caption{\textbf{Simulated M\o lmer-S\o rensen gate error under noisy drive.} We calculate and plot the gate infidelity as a function of the RPSD at the trap frequency $\nu$ (filled circles). We find that infidelity depends linearly in the RPSD at this specific frequency with a proportionality factor close to unity. This is shown by the linear fit to the numerical results (dashed line). The standard error is smaller than the marker size.}
    \label{fig:5}
\end{figure}

\emph{Full numerical simulation of two qubit gates.--}
We now combine our observation on both incoherent spin and motional dynamics to study the effect of fast phase noise on the MS gate. We use a similar numerical noise PSD where the noise overlaps with the trap frequency and a Hamiltonian with  Rabi frequency $\Omega=20$ kHz, $\nu=200$ kHz, and the Lamb-Dicke parameter $\eta= 0.15$. We solve the stochastic master equation for two qubits and their common motional state modeled as a quantum harmonic oscillator with 30 modes and find the gate fidelity as the overlap between the resulting density matrix and the target Bell-state. This is averaged over 1000 realizations of the same phase noise PSD. We repeat this for different noise amplitudes and plot the gate fidelity vs. the RPSD at the carrier multiplies with the gate time in Fig \ref{fig:5}. Once again, we find that this single parameter quantifies the performance of this two-qubit gate. We fit this trend with a linear model and find that the fidelity is proportional to the RPSD at the trap frequency with a proportionality factor near unity $1-F \simeq T\cdot RPSD(f=\nu)$. The nearly unity proportionality factor results from the fact that MS gates are long as compared with the inverse spectral width of typical noise features.

Similarly to our findings in a single ion, the mechanism behind this infidelity can be either due to incoherent spin pumping or heating of the ion. Since the MS gate is insensitive to the ion's temperature in the Lamb-Dicke regime, the leading source of infidelity is the incoherent spin pumping due to noise overlapping the carrier transition. This mechanism results in bit-flip errors during the gate that can further propagate. However, repeated operation of the MS gate as needed in a quantum circuit could eventually lead to considerable heating and larger gate errors. 


\section{Discussion}
In this paper, we study the effect of fast noise on the fidelity of quantum operations. For a broad class of single and two qubit operations including single qubit rotations with resonant transitions, off-resonant driving of multi-level single qubits, and two-qubits entangling gates, we identify a single parameter which quantifies the rate of errors or decoherence during the drive. This parameter is the noise Rabi PSD at the relevant frequencies, that is the Rabi frequency for resonant rotations and the detuning from the carrier transition for off-resonant drives and the MS gate. 

We find that the infidelity of single qubit rotations via resonant drive is set by the spectral overlap of the Rabi frequency and the noise PSD. For off-resonant drives or operations on sideband transitions, we identify two main channels for errors. The first one is due to a incoherent spin-pumping on the carrier transition and the second is coupling to higher excited motional modes, which takes the system out of the Hilbert space of exchanging a single motional quanta and results in effective heating. The latter becomes more significant with more drive cycles as the motional excitation accumulates.
We show that the heating mechanism has a negligible contribution to errors in a single operation of a two-qubit gate, whereas the incoherent spin pumping plays a pivotal role.

Assuming a trapped-ions quantum register with a carrier Rabi frequency of $\Omega=2\pi \times$100 kHz and a motional mode with a Lamb-Dicke parameter $\eta=0.05$, a MS gate using this mode will last 100 $\mu$s. Our findings indicate that in order to achieve an error below $10^{-4}$ we require the phase noise overlapping the carrier, in  terms of RPSD to be under $ \thicksim 1 [\frac{Hz^2}{Hz}]$. In terms of dBc this represents a requirement of -100 dBc/Hz on the RPSD in this frequency range. For weak noise the RPSD is proportional to the phase PSD and we have the -100 dBc/Hz requirement on phase PSD as well. This is not necessarily an easy goal to achieve. As an example, using a narrow linewidth laser operating on an optical qubit, with 100 kHz wide servo bump overlapping with the trap frequency, no more than $\simeq 10^{-5}$ of the laser intensity can be contained within this servo bump.

The fact that the gate error is well approximated by the RPSD at a single frequency results from the fact that typically the gate time is longer than the correlation time of the noise at these frequencies. In our case, the servo bump is spectrally wider than the Fourier width of the gate. This is not always true in trapped ion systems, and in particular we can find short single qubit $\pi$-times that deviate from this assumption. In these cases a proper overlap integral is necessary to estimate the gate error \cite{Chen2021}. 

Our analysis indicates that for full quantum control of operations at the high fidelity frontier it is important to characterise and study such fast noise mechanisms in any system. We note that our analysis is valid for any noise at the relevant spectral window which may overlap characteristic Hamiltonian energy scale and that the intricate dynamics coupling spin and motion under a noisy drive in the MS gate are relevant to any system where a bosonic mode mediates interaction between two spin-like qubits, such as gates between superconducting qubits mediated by a resonator. 
Our findings, analysis and detailed numerical calculations may thus guide the design and further improvement of quantum hardware and tailored quantum gates.

\section{acknowledgements}
This work was supported by the Israeli Science Foundation and the Goldring Family Foundation. We thank Yotam Shapira and Tom Manovitz for fruitful discussions. 

\bibliographystyle{latexmkrc}
\bibliography{phasenoise}

\begin{thebibliography}{34}%
\makeatletter
\providecommand \@ifxundefined [1]{%
 \@ifx{#1\undefined}
}%
\providecommand \@ifnum [1]{%
 \ifnum #1\expandafter \@firstoftwo
 \else \expandafter \@secondoftwo
 \fi
}%
\providecommand \@ifx [1]{%
 \ifx #1\expandafter \@firstoftwo
 \else \expandafter \@secondoftwo
 \fi
}%
\providecommand \natexlab [1]{#1}%
\providecommand \enquote  [1]{``#1''}%
\providecommand \bibnamefont  [1]{#1}%
\providecommand \bibfnamefont [1]{#1}%
\providecommand \citenamefont [1]{#1}%
\providecommand \href@noop [0]{\@secondoftwo}%
\providecommand \href [0]{\begingroup \@sanitize@url \@href}%
\providecommand \@href[1]{\@@startlink{#1}\@@href}%
\providecommand \@@href[1]{\endgroup#1\@@endlink}%
\providecommand \@sanitize@url [0]{\catcode `\\12\catcode `\$12\catcode
  `\&12\catcode `\#12\catcode `\^12\catcode `\_12\catcode `\%12\relax}%
\providecommand \@@startlink[1]{}%
\providecommand \@@endlink[0]{}%
\providecommand \url  [0]{\begingroup\@sanitize@url \@url }%
\providecommand \@url [1]{\endgroup\@href {#1}{\urlprefix }}%
\providecommand \urlprefix  [0]{URL }%
\providecommand \Eprint [0]{\href }%
\providecommand \doibase [0]{https://doi.org/}%
\providecommand \selectlanguage [0]{\@gobble}%
\providecommand \bibinfo  [0]{\@secondoftwo}%
\providecommand \bibfield  [0]{\@secondoftwo}%
\providecommand \translation [1]{[#1]}%
\providecommand \BibitemOpen [0]{}%
\providecommand \bibitemStop [0]{}%
\providecommand \bibitemNoStop [0]{.\EOS\space}%
\providecommand \EOS [0]{\spacefactor3000\relax}%
\providecommand \BibitemShut  [1]{\csname bibitem#1\endcsname}%
\let\auto@bib@innerbib\@empty
\bibitem [{\citenamefont {Benhelm}\ \emph {et~al.}(2008)\citenamefont
  {Benhelm}, \citenamefont {Kirchmair}, \citenamefont {Roos},\ and\
  \citenamefont {Blatt}}]{benhelm2008towards}%
  \BibitemOpen
  \bibfield  {author} {\bibinfo {author} {\bibfnamefont {J.}~\bibnamefont
  {Benhelm}}, \bibinfo {author} {\bibfnamefont {G.}~\bibnamefont {Kirchmair}},
  \bibinfo {author} {\bibfnamefont {C.~F.}\ \bibnamefont {Roos}},\ and\
  \bibinfo {author} {\bibfnamefont {R.}~\bibnamefont {Blatt}},\ }\bibfield
  {title} {\emph {\bibinfo {title} {Towards fault-tolerant quantum computing
  with trapped ions}},\ }\href {https://doi.org/10.1038/nphys961} {\bibfield
  {journal} {\bibinfo  {journal} {Nat. Phys.}\ }\textbf {\bibinfo {volume}
  {4}},\ \bibinfo {pages} {463} (\bibinfo {year} {2008})}\BibitemShut {NoStop}%
\bibitem [{\citenamefont {Schindler}\ \emph {et~al.}(2013)\citenamefont
  {Schindler}, \citenamefont {Nigg}, \citenamefont {Monz}, \citenamefont
  {Barreiro}, \citenamefont {Martinez}, \citenamefont {Wang}, \citenamefont
  {Quint}, \citenamefont {Brandl}, \citenamefont {Nebendahl}, \citenamefont
  {Roos} \emph {et~al.}}]{schindler2013quantum}%
  \BibitemOpen
  \bibfield  {author} {\bibinfo {author} {\bibfnamefont {P.}~\bibnamefont
  {Schindler}}, \bibinfo {author} {\bibfnamefont {D.}~\bibnamefont {Nigg}},
  \bibinfo {author} {\bibfnamefont {T.}~\bibnamefont {Monz}}, \bibinfo {author}
  {\bibfnamefont {J.~T.}\ \bibnamefont {Barreiro}}, \bibinfo {author}
  {\bibfnamefont {E.}~\bibnamefont {Martinez}}, \bibinfo {author}
  {\bibfnamefont {S.~X.}\ \bibnamefont {Wang}}, \bibinfo {author}
  {\bibfnamefont {S.}~\bibnamefont {Quint}}, \bibinfo {author} {\bibfnamefont
  {M.~F.}\ \bibnamefont {Brandl}}, \bibinfo {author} {\bibfnamefont
  {V.}~\bibnamefont {Nebendahl}}, \bibinfo {author} {\bibfnamefont {C.~F.}\
  \bibnamefont {Roos}}, \emph {et~al.},\ }\bibfield  {title} {\emph {\bibinfo
  {title} {A quantum information processor with trapped ions}},\ }\href
  {https://iopscience.iop.org/article/10.1088/1367-2630/15/12/123012/meta}
  {\bibfield  {journal} {\bibinfo  {journal} {New J. Phys.}\ }\textbf {\bibinfo
  {volume} {15}},\ \bibinfo {pages} {123012} (\bibinfo {year}
  {2013})}\BibitemShut {NoStop}%
\bibitem [{\citenamefont {Ban}(1998)}]{ban1998photon}%
  \BibitemOpen
  \bibfield  {author} {\bibinfo {author} {\bibfnamefont {M.}~\bibnamefont
  {Ban}},\ }\bibfield  {title} {\emph {\bibinfo {title} {Photon-echo technique
  for reducing the decoherence of a quantum bit}},\ }\href
  {https://doi.org/10.1080/09500349808231241} {\bibfield  {journal} {\bibinfo
  {journal} {Journal of Modern Optics}\ }\textbf {\bibinfo {volume} {45}},\
  \bibinfo {pages} {2315} (\bibinfo {year} {1998})}\BibitemShut {NoStop}%
\bibitem [{\citenamefont {Viola}\ \emph {et~al.}(1999)\citenamefont {Viola},
  \citenamefont {Knill},\ and\ \citenamefont {Lloyd}}]{PhysRevLett.82.2417}%
  \BibitemOpen
  \bibfield  {author} {\bibinfo {author} {\bibfnamefont {L.}~\bibnamefont
  {Viola}}, \bibinfo {author} {\bibfnamefont {E.}~\bibnamefont {Knill}},\ and\
  \bibinfo {author} {\bibfnamefont {S.}~\bibnamefont {Lloyd}},\ }\bibfield
  {title} {\emph {\bibinfo {title} {Dynamical decoupling of open quantum
  systems}},\ }\href {https://doi.org/10.1103/PhysRevLett.82.2417} {\bibfield
  {journal} {\bibinfo  {journal} {Phys. Rev. Lett.}\ }\textbf {\bibinfo
  {volume} {82}},\ \bibinfo {pages} {2417} (\bibinfo {year}
  {1999})}\BibitemShut {NoStop}%
\bibitem [{\citenamefont {Vitali}\ and\ \citenamefont
  {Tombesi}(1999)}]{PhysRevA.59.4178}%
  \BibitemOpen
  \bibfield  {author} {\bibinfo {author} {\bibfnamefont {D.}~\bibnamefont
  {Vitali}}\ and\ \bibinfo {author} {\bibfnamefont {P.}~\bibnamefont
  {Tombesi}},\ }\bibfield  {title} {\emph {\bibinfo {title} {Using parity kicks
  for decoherence control}},\ }\href {https://doi.org/10.1103/PhysRevA.59.4178}
  {\bibfield  {journal} {\bibinfo  {journal} {Phys. Rev. A}\ }\textbf {\bibinfo
  {volume} {59}},\ \bibinfo {pages} {4178} (\bibinfo {year}
  {1999})}\BibitemShut {NoStop}%
\bibitem [{\citenamefont {Viola}\ and\ \citenamefont
  {Lloyd}(1998)}]{PhysRevA.58.2733}%
  \BibitemOpen
  \bibfield  {author} {\bibinfo {author} {\bibfnamefont {L.}~\bibnamefont
  {Viola}}\ and\ \bibinfo {author} {\bibfnamefont {S.}~\bibnamefont {Lloyd}},\
  }\bibfield  {title} {\emph {\bibinfo {title} {Dynamical suppression of
  decoherence in two-state quantum systems}},\ }\href
  {https://doi.org/10.1103/PhysRevA.58.2733} {\bibfield  {journal} {\bibinfo
  {journal} {Phys. Rev. A}\ }\textbf {\bibinfo {volume} {58}},\ \bibinfo
  {pages} {2733} (\bibinfo {year} {1998})}\BibitemShut {NoStop}%
\bibitem [{\citenamefont {Carr}\ and\ \citenamefont
  {Purcell}(1954)}]{PhysRev.94.630}%
  \BibitemOpen
  \bibfield  {author} {\bibinfo {author} {\bibfnamefont {H.~Y.}\ \bibnamefont
  {Carr}}\ and\ \bibinfo {author} {\bibfnamefont {E.~M.}\ \bibnamefont
  {Purcell}},\ }\bibfield  {title} {\emph {\bibinfo {title} {Effects of
  diffusion on free precession in nuclear magnetic resonance experiments}},\
  }\href {https://doi.org/10.1103/PhysRev.94.630} {\bibfield  {journal}
  {\bibinfo  {journal} {Phys. Rev.}\ }\textbf {\bibinfo {volume} {94}},\
  \bibinfo {pages} {630} (\bibinfo {year} {1954})}\BibitemShut {NoStop}%
\bibitem [{\citenamefont {Meiboom}\ and\ \citenamefont
  {Gill}(1958)}]{meiboom1958modified}%
  \BibitemOpen
  \bibfield  {author} {\bibinfo {author} {\bibfnamefont {S.}~\bibnamefont
  {Meiboom}}\ and\ \bibinfo {author} {\bibfnamefont {D.}~\bibnamefont {Gill}},\
  }\bibfield  {title} {\emph {\bibinfo {title} {Modified spin-echo method for
  measuring nuclear relaxation times}},\ }\href@noop {} {\bibfield  {journal}
  {\bibinfo  {journal} {Review of scientific instruments}\ }\textbf {\bibinfo
  {volume} {29}},\ \bibinfo {pages} {688} (\bibinfo {year} {1958})}\BibitemShut
  {NoStop}%
\bibitem [{\citenamefont {Manovitz}\ \emph {et~al.}(2017)\citenamefont
  {Manovitz}, \citenamefont {Rotem}, \citenamefont {Shaniv}, \citenamefont
  {Cohen}, \citenamefont {Shapira}, \citenamefont {Akerman}, \citenamefont
  {Retzker},\ and\ \citenamefont {Ozeri}}]{PhysRevLett.119.220505}%
  \BibitemOpen
  \bibfield  {author} {\bibinfo {author} {\bibfnamefont {T.}~\bibnamefont
  {Manovitz}}, \bibinfo {author} {\bibfnamefont {A.}~\bibnamefont {Rotem}},
  \bibinfo {author} {\bibfnamefont {R.}~\bibnamefont {Shaniv}}, \bibinfo
  {author} {\bibfnamefont {I.}~\bibnamefont {Cohen}}, \bibinfo {author}
  {\bibfnamefont {Y.}~\bibnamefont {Shapira}}, \bibinfo {author} {\bibfnamefont
  {N.}~\bibnamefont {Akerman}}, \bibinfo {author} {\bibfnamefont
  {A.}~\bibnamefont {Retzker}},\ and\ \bibinfo {author} {\bibfnamefont
  {R.}~\bibnamefont {Ozeri}},\ }\bibfield  {title} {\emph {\bibinfo {title}
  {Fast dynamical decoupling of the m\o{}lmer-s\o{}rensen entangling gate}},\
  }\href {https://doi.org/10.1103/PhysRevLett.119.220505} {\bibfield  {journal}
  {\bibinfo  {journal} {Phys. Rev. Lett.}\ }\textbf {\bibinfo {volume} {119}},\
  \bibinfo {pages} {220505} (\bibinfo {year} {2017})}\BibitemShut {NoStop}%
\bibitem [{\citenamefont {Lidar}\ \emph {et~al.}(1998)\citenamefont {Lidar},
  \citenamefont {Chuang},\ and\ \citenamefont {Whaley}}]{lidar1998decoherence}%
  \BibitemOpen
  \bibfield  {author} {\bibinfo {author} {\bibfnamefont {D.~A.}\ \bibnamefont
  {Lidar}}, \bibinfo {author} {\bibfnamefont {I.~L.}\ \bibnamefont {Chuang}},\
  and\ \bibinfo {author} {\bibfnamefont {K.~B.}\ \bibnamefont {Whaley}},\
  }\bibfield  {title} {\emph {\bibinfo {title} {Decoherence-free subspaces for
  quantum computation}},\ }\href
  {https://journals.aps.org/prl/abstract/10.1103/PhysRevLett.81.2594}
  {\bibfield  {journal} {\bibinfo  {journal} {Phys. Rev. Lett.}\ }\textbf
  {\bibinfo {volume} {81}},\ \bibinfo {pages} {2594} (\bibinfo {year}
  {1998})}\BibitemShut {NoStop}%
\bibitem [{\citenamefont {Barenco}\ \emph {et~al.}(1997)\citenamefont
  {Barenco}, \citenamefont {Berthiaume}, \citenamefont {Deutsch}, \citenamefont
  {Ekert}, \citenamefont {Jozsa},\ and\ \citenamefont
  {Macchiavello}}]{barenco1997stabilization}%
  \BibitemOpen
  \bibfield  {author} {\bibinfo {author} {\bibfnamefont {A.}~\bibnamefont
  {Barenco}}, \bibinfo {author} {\bibfnamefont {A.}~\bibnamefont {Berthiaume}},
  \bibinfo {author} {\bibfnamefont {D.}~\bibnamefont {Deutsch}}, \bibinfo
  {author} {\bibfnamefont {A.}~\bibnamefont {Ekert}}, \bibinfo {author}
  {\bibfnamefont {R.}~\bibnamefont {Jozsa}},\ and\ \bibinfo {author}
  {\bibfnamefont {C.}~\bibnamefont {Macchiavello}},\ }\bibfield  {title} {\emph
  {\bibinfo {title} {Stabilization of quantum computations by
  symmetrization}},\ }\href
  {https://epubs.siam.org/doi/abs/10.1137/S0097539796302452?casa_token=WaO46UW8ftQAAAAA:8zQsb_sAU4VwGRxg9OxPYfybxTOngkIdId8KAF1_Nq_4VpQLt99lqPtWS8A2GeKhryWem1ot9Q}
  {\bibfield  {journal} {\bibinfo  {journal} {SIAM Journal on Computing}\
  }\textbf {\bibinfo {volume} {26}},\ \bibinfo {pages} {1541} (\bibinfo {year}
  {1997})}\BibitemShut {NoStop}%
\bibitem [{\citenamefont {Zanardi}(2000)}]{PhysRevA.63.012301}%
  \BibitemOpen
  \bibfield  {author} {\bibinfo {author} {\bibfnamefont {P.}~\bibnamefont
  {Zanardi}},\ }\bibfield  {title} {\emph {\bibinfo {title} {Stabilizing
  quantum information}},\ }\href {https://doi.org/10.1103/PhysRevA.63.012301}
  {\bibfield  {journal} {\bibinfo  {journal} {Phys. Rev. A}\ }\textbf {\bibinfo
  {volume} {63}},\ \bibinfo {pages} {012301} (\bibinfo {year}
  {2000})}\BibitemShut {NoStop}%
\bibitem [{\citenamefont {Duan}\ and\ \citenamefont
  {Guo}(1997)}]{PhysRevLett.79.1953}%
  \BibitemOpen
  \bibfield  {author} {\bibinfo {author} {\bibfnamefont {L.-M.}\ \bibnamefont
  {Duan}}\ and\ \bibinfo {author} {\bibfnamefont {G.-C.}\ \bibnamefont {Guo}},\
  }\bibfield  {title} {\emph {\bibinfo {title} {Preserving coherence in quantum
  computation by pairing quantum bits}},\ }\href
  {https://doi.org/10.1103/PhysRevLett.79.1953} {\bibfield  {journal} {\bibinfo
   {journal} {Phys. Rev. Lett.}\ }\textbf {\bibinfo {volume} {79}},\ \bibinfo
  {pages} {1953} (\bibinfo {year} {1997})}\BibitemShut {NoStop}%
\bibitem [{\citenamefont {Zanardi}\ and\ \citenamefont
  {Rasetti}(1997)}]{PhysRevLett.79.3306}%
  \BibitemOpen
  \bibfield  {author} {\bibinfo {author} {\bibfnamefont {P.}~\bibnamefont
  {Zanardi}}\ and\ \bibinfo {author} {\bibfnamefont {M.}~\bibnamefont
  {Rasetti}},\ }\bibfield  {title} {\emph {\bibinfo {title} {Noiseless quantum
  codes}},\ }\href {https://doi.org/10.1103/PhysRevLett.79.3306} {\bibfield
  {journal} {\bibinfo  {journal} {Phys. Rev. Lett.}\ }\textbf {\bibinfo
  {volume} {79}},\ \bibinfo {pages} {3306} (\bibinfo {year}
  {1997})}\BibitemShut {NoStop}%
\bibitem [{\citenamefont {Ogata}\ \emph {et~al.}(2010)\citenamefont {Ogata}
  \emph {et~al.}}]{ogata2010modern}%
  \BibitemOpen
  \bibfield  {author} {\bibinfo {author} {\bibfnamefont {K.}~\bibnamefont
  {Ogata}} \emph {et~al.},\ }\href@noop {} {\emph {\bibinfo {title} {Modern
  control engineering}}},\ Vol.~\bibinfo {volume} {5}\ (\bibinfo  {publisher}
  {Prentice hall Upper Saddle River, NJ},\ \bibinfo {year} {2010})\BibitemShut
  {NoStop}%
\bibitem [{\citenamefont {Akerman}\ \emph {et~al.}(2015)\citenamefont
  {Akerman}, \citenamefont {Navon}, \citenamefont {Kotler}, \citenamefont
  {Glickman},\ and\ \citenamefont {Ozeri}}]{Akerman_2015}%
  \BibitemOpen
  \bibfield  {author} {\bibinfo {author} {\bibfnamefont {N.}~\bibnamefont
  {Akerman}}, \bibinfo {author} {\bibfnamefont {N.}~\bibnamefont {Navon}},
  \bibinfo {author} {\bibfnamefont {S.}~\bibnamefont {Kotler}}, \bibinfo
  {author} {\bibfnamefont {Y.}~\bibnamefont {Glickman}},\ and\ \bibinfo
  {author} {\bibfnamefont {R.}~\bibnamefont {Ozeri}},\ }\bibfield  {title}
  {\emph {\bibinfo {title} {Universal gate-set for trapped-ion qubits using a
  narrow linewidth diode laser}},\ }\href
  {https://doi.org/10.1088/1367-2630/17/11/113060} {\bibfield  {journal}
  {\bibinfo  {journal} {New J. Phys.}\ }\textbf {\bibinfo {volume} {17}},\
  \bibinfo {pages} {113060} (\bibinfo {year} {2015})}\BibitemShut {NoStop}%
\bibitem [{\citenamefont {Kotler}\ \emph {et~al.}(2013)\citenamefont {Kotler},
  \citenamefont {Akerman}, \citenamefont {Glickman},\ and\ \citenamefont
  {Ozeri}}]{PhysRevLett.110.110503}%
  \BibitemOpen
  \bibfield  {author} {\bibinfo {author} {\bibfnamefont {S.}~\bibnamefont
  {Kotler}}, \bibinfo {author} {\bibfnamefont {N.}~\bibnamefont {Akerman}},
  \bibinfo {author} {\bibfnamefont {Y.}~\bibnamefont {Glickman}},\ and\
  \bibinfo {author} {\bibfnamefont {R.}~\bibnamefont {Ozeri}},\ }\bibfield
  {title} {\emph {\bibinfo {title} {Nonlinear single-spin spectrum analyzer}},\
  }\href {https://doi.org/10.1103/PhysRevLett.110.110503} {\bibfield  {journal}
  {\bibinfo  {journal} {Phys. Rev. Lett.}\ }\textbf {\bibinfo {volume} {110}},\
  \bibinfo {pages} {110503} (\bibinfo {year} {2013})}\BibitemShut {NoStop}%
\bibitem [{\citenamefont {Fanciulli}(2009)}]{fanciulli2009electron}%
  \BibitemOpen
  \bibfield  {author} {\bibinfo {author} {\bibfnamefont {M.}~\bibnamefont
  {Fanciulli}},\ }\href@noop {} {\emph {\bibinfo {title} {Electron spin
  resonance and related phenomena in low-dimensional structures}}},\ Vol.\
  \bibinfo {volume} {115}\ (\bibinfo  {publisher} {Springer Science \& Business
  Media},\ \bibinfo {year} {2009})\BibitemShut {NoStop}%
\bibitem [{\citenamefont {Hall}\ \emph {et~al.}(2009)\citenamefont {Hall},
  \citenamefont {Cole}, \citenamefont {Hill},\ and\ \citenamefont
  {Hollenberg}}]{PhysRevLett.103.220802}%
  \BibitemOpen
  \bibfield  {author} {\bibinfo {author} {\bibfnamefont {L.~T.}\ \bibnamefont
  {Hall}}, \bibinfo {author} {\bibfnamefont {J.~H.}\ \bibnamefont {Cole}},
  \bibinfo {author} {\bibfnamefont {C.~D.}\ \bibnamefont {Hill}},\ and\
  \bibinfo {author} {\bibfnamefont {L.~C.~L.}\ \bibnamefont {Hollenberg}},\
  }\bibfield  {title} {\emph {\bibinfo {title} {Sensing of fluctuating
  nanoscale magnetic fields using nitrogen-vacancy centers in diamond}},\
  }\href {https://doi.org/10.1103/PhysRevLett.103.220802} {\bibfield  {journal}
  {\bibinfo  {journal} {Phys. Rev. Lett.}\ }\textbf {\bibinfo {volume} {103}},\
  \bibinfo {pages} {220802} (\bibinfo {year} {2009})}\BibitemShut {NoStop}%
\bibitem [{\citenamefont {Cywi\ifmmode~\acute{n}\else \'{n}\fi{}ski}\ \emph
  {et~al.}(2008)\citenamefont {Cywi\ifmmode~\acute{n}\else \'{n}\fi{}ski},
  \citenamefont {Lutchyn}, \citenamefont {Nave},\ and\ \citenamefont
  {Das~Sarma}}]{PhysRevB.77.174509}%
  \BibitemOpen
  \bibfield  {author} {\bibinfo {author} {\bibfnamefont {L.}~\bibnamefont
  {Cywi\ifmmode~\acute{n}\else \'{n}\fi{}ski}}, \bibinfo {author}
  {\bibfnamefont {R.~M.}\ \bibnamefont {Lutchyn}}, \bibinfo {author}
  {\bibfnamefont {C.~P.}\ \bibnamefont {Nave}},\ and\ \bibinfo {author}
  {\bibfnamefont {S.}~\bibnamefont {Das~Sarma}},\ }\bibfield  {title} {\emph
  {\bibinfo {title} {How to enhance dephasing time in superconducting
  qubits}},\ }\href {https://doi.org/10.1103/PhysRevB.77.174509} {\bibfield
  {journal} {\bibinfo  {journal} {Phys. Rev. B}\ }\textbf {\bibinfo {volume}
  {77}},\ \bibinfo {pages} {174509} (\bibinfo {year} {2008})}\BibitemShut
  {NoStop}%
\bibitem [{\citenamefont {Lasič}\ \emph {et~al.}(2006)\citenamefont {Lasič},
  \citenamefont {Stepišnik},\ and\ \citenamefont {Mohorič}}]{LASIC2006208}%
  \BibitemOpen
  \bibfield  {author} {\bibinfo {author} {\bibfnamefont {S.}~\bibnamefont
  {Lasič}}, \bibinfo {author} {\bibfnamefont {J.}~\bibnamefont {Stepišnik}},\
  and\ \bibinfo {author} {\bibfnamefont {A.}~\bibnamefont {Mohorič}},\
  }\bibfield  {title} {\emph {\bibinfo {title} {Displacement power spectrum
  measurement by cpmg in constant gradient}},\ }\href
  {https://doi.org/https://doi.org/10.1016/j.jmr.2006.06.030} {\bibfield
  {journal} {\bibinfo  {journal} {Journal of Magnetic Resonance}\ }\textbf
  {\bibinfo {volume} {182}},\ \bibinfo {pages} {208} (\bibinfo {year}
  {2006})}\BibitemShut {NoStop}%
\bibitem [{\citenamefont {Day}\ \emph {et~al.}(2022)\citenamefont {Day},
  \citenamefont {Low}, \citenamefont {White}, \citenamefont {Islam},\ and\
  \citenamefont {Senko}}]{Day2022}%
  \BibitemOpen
  \bibfield  {author} {\bibinfo {author} {\bibfnamefont {M.~L.}\ \bibnamefont
  {Day}}, \bibinfo {author} {\bibfnamefont {P.~J.}\ \bibnamefont {Low}},
  \bibinfo {author} {\bibfnamefont {B.}~\bibnamefont {White}}, \bibinfo
  {author} {\bibfnamefont {R.}~\bibnamefont {Islam}},\ and\ \bibinfo {author}
  {\bibfnamefont {C.}~\bibnamefont {Senko}},\ }\bibfield  {title} {\emph
  {\bibinfo {title} {Limits on atomic qubit control from laser noise}},\ }\href
  {https://doi.org/10.1038/s41534-022-00586-4} {\bibfield  {journal} {\bibinfo
  {journal} {npj Quantum Information}\ }\textbf {\bibinfo {volume} {8}}
  (\bibinfo {year} {2022})}\BibitemShut {NoStop}%
\bibitem [{\citenamefont {Paschotta}\ \emph {et~al.}(2017)\citenamefont
  {Paschotta}, \citenamefont {Telle},\ and\ \citenamefont {Keller}}]{inbook}%
  \BibitemOpen
  \bibfield  {author} {\bibinfo {author} {\bibfnamefont {R.}~\bibnamefont
  {Paschotta}}, \bibinfo {author} {\bibfnamefont {H.}~\bibnamefont {Telle}},\
  and\ \bibinfo {author} {\bibfnamefont {U.}~\bibnamefont {Keller}},\ }\bibinfo
  {title} {Noise of solid-state lasers}\ (\bibinfo {year} {2017})\ pp.\
  \bibinfo {pages} {473--510}\BibitemShut {NoStop}%
\bibitem [{\citenamefont {Langbein}(2004)}]{langbein2004noise}%
  \BibitemOpen
  \bibfield  {author} {\bibinfo {author} {\bibfnamefont {J.}~\bibnamefont
  {Langbein}},\ }\bibfield  {title} {\emph {\bibinfo {title} {Noise in
  two-color electronic distance meter measurements revisited}},\ }\href@noop {}
  {\bibfield  {journal} {\bibinfo  {journal} {Journal of Geophysical Research:
  Solid Earth}\ }\textbf {\bibinfo {volume} {109}} (\bibinfo {year}
  {2004})}\BibitemShut {NoStop}%
\bibitem [{\citenamefont {Xu}(2019)}]{xu2019easy}%
  \BibitemOpen
  \bibfield  {author} {\bibinfo {author} {\bibfnamefont {C.}~\bibnamefont
  {Xu}},\ }\bibfield  {title} {\emph {\bibinfo {title} {An easy algorithm to
  generate colored noise sequences}},\ }\href {https://doi.org/Chang Xu 2019 AJ
  157 127} {\bibfield  {journal} {\bibinfo  {journal} {The Astronomical
  Journal}\ }\textbf {\bibinfo {volume} {157}},\ \bibinfo {pages} {127}
  (\bibinfo {year} {2019})}\BibitemShut {NoStop}%
\bibitem [{\citenamefont {Johansson}\ \emph {et~al.}(2012)\citenamefont
  {Johansson}, \citenamefont {Nation},\ and\ \citenamefont
  {Nori}}]{johansson2012qutip}%
  \BibitemOpen
  \bibfield  {author} {\bibinfo {author} {\bibfnamefont {J.~R.}\ \bibnamefont
  {Johansson}}, \bibinfo {author} {\bibfnamefont {P.~D.}\ \bibnamefont
  {Nation}},\ and\ \bibinfo {author} {\bibfnamefont {F.}~\bibnamefont {Nori}},\
  }\bibfield  {title} {\emph {\bibinfo {title} {Qutip: An open-source python
  framework for the dynamics of open quantum systems}},\ }\href@noop {}
  {\bibfield  {journal} {\bibinfo  {journal} {Computer Physics Communications}\
  }\textbf {\bibinfo {volume} {183}},\ \bibinfo {pages} {1760} (\bibinfo {year}
  {2012})}\BibitemShut {NoStop}%
\bibitem [{\citenamefont {Peleg}\ \emph {et~al.}()\citenamefont {Peleg},
  \citenamefont {Akerman}, \citenamefont {Manovit}, \citenamefont {Alon},\ and\
  \citenamefont {Ozeri}}]{Peleg2019}%
  \BibitemOpen
  \bibfield  {author} {\bibinfo {author} {\bibfnamefont {L.}~\bibnamefont
  {Peleg}}, \bibinfo {author} {\bibfnamefont {N.}~\bibnamefont {Akerman}},
  \bibinfo {author} {\bibfnamefont {T.}~\bibnamefont {Manovit}}, \bibinfo
  {author} {\bibfnamefont {M.}~\bibnamefont {Alon}},\ and\ \bibinfo {author}
  {\bibfnamefont {R.}~\bibnamefont {Ozeri}},\ }\bibfield  {title} {\emph
  {\bibinfo {title} {Phase stability transfer across the optical domain using a
  commercial optical frequency comb system}},\ }\bibfield  {journal} {\bibinfo
  {journal} {arXiv.1905.05065}\ }\href
  {https://doi.org/https://doi.org/10.48550/arXiv.1905.05065}
  {https://doi.org/10.48550/arXiv.1905.05065}\BibitemShut {NoStop}%
\bibitem [{\citenamefont {Kotler}\ \emph {et~al.}(2011)\citenamefont {Kotler},
  \citenamefont {Akerman}, \citenamefont {Glickman}, \citenamefont {Keselman},\
  and\ \citenamefont {Ozeri}}]{Kotler2011}%
  \BibitemOpen
  \bibfield  {author} {\bibinfo {author} {\bibfnamefont {S.}~\bibnamefont
  {Kotler}}, \bibinfo {author} {\bibfnamefont {N.}~\bibnamefont {Akerman}},
  \bibinfo {author} {\bibfnamefont {Y.}~\bibnamefont {Glickman}}, \bibinfo
  {author} {\bibfnamefont {A.}~\bibnamefont {Keselman}},\ and\ \bibinfo
  {author} {\bibfnamefont {R.}~\bibnamefont {Ozeri}},\ }\bibfield  {title}
  {\emph {\bibinfo {title} {Single-ion quantum lock-in amplifier}},\ }\href
  {https://www.nature.com/articles/nature10010} {\bibfield  {journal} {\bibinfo
   {journal} {Nature}\ }\textbf {\bibinfo {volume} {473}},\ \bibinfo {pages}
  {61} (\bibinfo {year} {2011})}\BibitemShut {NoStop}%
\bibitem [{\citenamefont {Christian L~Degen}(2017)}]{DegenRMP2017}%
  \BibitemOpen
  \bibfield  {author} {\bibinfo {author} {\bibfnamefont {P.~C.}\ \bibnamefont
  {Christian L~Degen}, \bibfnamefont {Friedemann~Reinhard}},\ }\bibfield
  {title} {\emph {\bibinfo {title} {Quantum sensing}},\ }\href
  {https://journals.aps.org/rmp/abstract/10.1103/RevModPhys.89.035002}
  {\bibfield  {journal} {\bibinfo  {journal} {Rev. Mod. Phys.}\ }\textbf
  {\bibinfo {volume} {89}},\ \bibinfo {pages} {035002} (\bibinfo {year}
  {2017})}\BibitemShut {NoStop}%
\bibitem [{\citenamefont {Chen}\ \emph {et~al.}(2012)\citenamefont {Chen},
  \citenamefont {Bohnet}, \citenamefont {Weiner},\ and\ \citenamefont
  {Thompson}}]{Chen2021}%
  \BibitemOpen
  \bibfield  {author} {\bibinfo {author} {\bibfnamefont {Z.}~\bibnamefont
  {Chen}}, \bibinfo {author} {\bibfnamefont {J.~G.}\ \bibnamefont {Bohnet}},
  \bibinfo {author} {\bibfnamefont {J.~M.}\ \bibnamefont {Weiner}},\ and\
  \bibinfo {author} {\bibfnamefont {J.~K.}\ \bibnamefont {Thompson}},\
  }\bibfield  {title} {\emph {\bibinfo {title} {General formalism for
  evaluating the impact of phase noise on bloch vector rotations}},\ }\href
  {https://doi.org/10.1103/PhysRevA.86.032313} {\bibfield  {journal} {\bibinfo
  {journal} {Phys. Rev. A}\ }\textbf {\bibinfo {volume} {86}},\ \bibinfo
  {pages} {032313} (\bibinfo {year} {2012})}\BibitemShut {NoStop}%
\bibitem [{\citenamefont {S{\o}rensen}\ and\ \citenamefont
  {M{\o}lmer}(1999)}]{sorensen1999quantum}%
  \BibitemOpen
  \bibfield  {author} {\bibinfo {author} {\bibfnamefont {A.}~\bibnamefont
  {S{\o}rensen}}\ and\ \bibinfo {author} {\bibfnamefont {K.}~\bibnamefont
  {M{\o}lmer}},\ }\bibfield  {title} {\emph {\bibinfo {title} {Quantum
  computation with ions in thermal motion}},\ }\href
  {https://journals.aps.org/prl/abstract/10.1103/PhysRevLett.82.1971}
  {\bibfield  {journal} {\bibinfo  {journal} {Phys. Rev. Lett.}\ }\textbf
  {\bibinfo {volume} {82}},\ \bibinfo {pages} {1971} (\bibinfo {year}
  {1999})}\BibitemShut {NoStop}%
\bibitem [{\citenamefont {S{\o}rensen}\ and\ \citenamefont
  {M{\o}lmer}(2000)}]{sorensen2000entanglement}%
  \BibitemOpen
  \bibfield  {author} {\bibinfo {author} {\bibfnamefont {A.}~\bibnamefont
  {S{\o}rensen}}\ and\ \bibinfo {author} {\bibfnamefont {K.}~\bibnamefont
  {M{\o}lmer}},\ }\bibfield  {title} {\emph {\bibinfo {title} {Entanglement and
  quantum computation with ions in thermal motion}},\ }\href
  {https://journals.aps.org/pra/abstract/10.1103/PhysRevA.62.022311} {\bibfield
   {journal} {\bibinfo  {journal} {Phys. Rev. A}\ }\textbf {\bibinfo {volume}
  {62}},\ \bibinfo {pages} {022311} (\bibinfo {year} {2000})}\BibitemShut
  {NoStop}%
\bibitem [{\citenamefont {Meekhof}\ \emph {et~al.}(1996)\citenamefont
  {Meekhof}, \citenamefont {Monroe}, \citenamefont {King}, \citenamefont
  {Itano},\ and\ \citenamefont {Wineland}}]{meekhof1996}%
  \BibitemOpen
  \bibfield  {author} {\bibinfo {author} {\bibfnamefont {D.}~\bibnamefont
  {Meekhof}}, \bibinfo {author} {\bibfnamefont {C.}~\bibnamefont {Monroe}},
  \bibinfo {author} {\bibfnamefont {B.}~\bibnamefont {King}}, \bibinfo {author}
  {\bibfnamefont {W.~M.}\ \bibnamefont {Itano}},\ and\ \bibinfo {author}
  {\bibfnamefont {D.~J.}\ \bibnamefont {Wineland}},\ }\bibfield  {title} {\emph
  {\bibinfo {title} {Generation of nonclassical motional states of a trapped
  atom}},\ }\href
  {https://journals.aps.org/prl/abstract/10.1103/PhysRevLett.76.1796}
  {\bibfield  {journal} {\bibinfo  {journal} {Phys. Rev. Lett.}\ }\textbf
  {\bibinfo {volume} {76}},\ \bibinfo {pages} {1796} (\bibinfo {year}
  {1996})}\BibitemShut {NoStop}%
\bibitem [{\citenamefont {Cai}\ \emph {et~al.}(2021)\citenamefont {Cai},
  \citenamefont {Liu}, \citenamefont {Zhao}, \citenamefont {Wu}, \citenamefont
  {Mei}, \citenamefont {Jiang}, \citenamefont {He}, \citenamefont {Zhang},
  \citenamefont {Zhou},\ and\ \citenamefont {Duan}}]{cai2021}%
  \BibitemOpen
  \bibfield  {author} {\bibinfo {author} {\bibfnamefont {M.-L.}\ \bibnamefont
  {Cai}}, \bibinfo {author} {\bibfnamefont {Z.-D.}\ \bibnamefont {Liu}},
  \bibinfo {author} {\bibfnamefont {W.-D.}\ \bibnamefont {Zhao}}, \bibinfo
  {author} {\bibfnamefont {Y.-K.}\ \bibnamefont {Wu}}, \bibinfo {author}
  {\bibfnamefont {Q.-X.}\ \bibnamefont {Mei}}, \bibinfo {author} {\bibfnamefont
  {Y.}~\bibnamefont {Jiang}}, \bibinfo {author} {\bibfnamefont
  {L.}~\bibnamefont {He}}, \bibinfo {author} {\bibfnamefont {X.}~\bibnamefont
  {Zhang}}, \bibinfo {author} {\bibfnamefont {Z.-C.}\ \bibnamefont {Zhou}},\
  and\ \bibinfo {author} {\bibfnamefont {L.-M.}\ \bibnamefont {Duan}},\
  }\bibfield  {title} {\emph {\bibinfo {title} {Observation of a quantum phase
  transition in the quantum rabi model with a single trapped ion}},\ }\href
  {https://doi.org/Nat Commun 12, 1126 (2021)} {\bibfield  {journal} {\bibinfo
  {journal} {Nat. commun.}\ }\textbf {\bibinfo {volume} {12}},\ \bibinfo
  {pages} {1} (\bibinfo {year} {2021})}\BibitemShut {NoStop}%
\end{thebibliography}%
\end{document}